\documentclass[preprint]{revtex4}
\usepackage{graphicx}
\usepackage{soul}
\usepackage{ulem}
\usepackage{amssymb}
\newcommand{\comment}[1]{}
\def\ktresummation{$k_t$-resummation}
\def\ee{\end{equation}}
\def\be{\begin{equation}}
\def\eea{\end{eqnarray}}
\def\bea{\begin{eqnarray}}

\def\pp{$pp$}
\def\pbarp{${\bar p}p$}

\def\nsoft{{\bar n}_{soft}}

\def\ptmin{p_{tmin}}
\def\rs{\sqrt{s}}

\def\shat{{\hat s}}

\def\vecb{{\vec b}}

\def\nbar{{\bar n}}

\begin{document}


\title{Total and inelastic cross-sections at LHC at $\sqrt{s}=7 \ TeV$ and beyond}


\author{Andrea Achilli}
\email[]{andrea.achilli@fisica.unipg.it}
\affiliation{INFN \& Dipartimento di Fisica, Universita' di Perugia, Perugia, Italy }
\author{Rohini Godbole}
\email[]{rohini@cts.iisc.ernet.in}
\affiliation{Institute for Theoretical Physics and Spinoza Institute, Utrecht
  University, 3508 TD Utrecht, The Netherlands,}
  \affiliation{ Centre for High Energy Physics, Indian Institute of Science, Bangalore, 560 012, India}
\author{Agnes Grau}
\email[]{igrau@ugr.es}
\affiliation{Departamento de F\'\i sica Te\'orica y del Cosmos,Universidad de Granada, 18071 Granada, Spain}
\author{Giulia Pancheri}
\email[]{giulia.pancheri@lnf.infn.it}
\affiliation{Laboratori Nazionali di Frascati, INFN, Frascati,Italy}
\author{Olga Shekhovtsova}
\email[]{olga.shekhovtsova@ific.uv.es}
\affiliation{IFIC,Universitat de Valencia-CSIC, Apt. Correus 22085, E-46071, Valencia, Spain}
\author{Yogendra Srivastava}
\email[]{yogendra.srivastava@pg.infn.it}
\affiliation{INFN \& Dipartimento di Fisica, Universita' di Perugia, Perugia, Italy }

\date{\today}

\begin{abstract}
We  discuss  expectations   for the total and inelastic cross-sections
at  
  LHC CM energies $\sqrt{s}\ =\ 7\ TeV$ { and $ 14\ TeV$}   obtained in an eikonal minijet model augmented by soft gluon $k_t$-resummation, 
 which we describe  in some detail.  We   present a band of predictions which encompass   recent LHC data and    suggest  that the inelastic cross-section described by two channel eikonal models      include only  uncorrelated  processes. We  show that this  interpretation of the model is supported by  the LHC data .
 \end{abstract}


\maketitle

\section{Introduction \label{Intro}}
Models for the high energy behaviour of the total cross-section hold a large 
interest since  Heisenberg's early attempts to describe the production of 
mesons through a shock wave mechanism \cite{Heisenberg:1952zz}.  The  recent 
measurement  of the  inelastic cross-section  at c.m. energy of 7 TeV by the 
ATLAS \cite{Aad:2011eu}    collaboration,  together with preliminary CMS 
results~\cite{cmsnote} 
and the expected results from other LHC groups, have renewed the interest in the subject \cite{Achilli:2011sw,Block:2011uy,Goulianos:2011pk,Jenkovszky:2011hu}.  In addition, very recently, the TOTEM collaboration has  released    values for  the total, the elastic and the inelastic  cross-sections
\cite{Latino:arXiv1110.1008} and the  Auger experiment  has given  a  measurement of the $p-air$ cross-section at $\sqrt{s}=57\ TeV$ \cite{:2011pe}.

The  interest arises because total  cross-sections access  the large distance regime of particle interactions, i.e. a region where  a QCD description is still lacking.  In  this paper, we present  estimates for  the total and the inelastic 
cross-sections at the  LHC  energies, $\sqrt{s}\ =\ 7\ $ and $14\ TeV$,
based on an eikonal minijet model  augmented by soft-gluon 
resummation which we have developed over the 
years \cite{Corsetti:1996wg,Grau:1999em,Godbole:2004kx}. Our present aim is not to give newer fits to the data, but to  use the model to further develop an understanding of the underlying large distance dynamics. This has led us to an  interesting result, discussed in this paper, namely that the expression for the inelastic cross-section obtained from simple two channel eikonal models only describes the contribution of uncorrelated scattering processes. 
 
 The eikonal model for high energy scattering  \cite{Glauber:1959fg,Glauber:1970jm} provides a framework which incorporates unitarity in the description of scattering processes and  can be used  to describe both the total and the inelastic cross-section.    Following up on the earliest discussions of the role of  low $p_t$ jets (aka mini-jets)
in affecting the energy dependence of total cross-sections \cite{Cline:1973kv,Gaisser:1984pg,Pancheri:1986qg}, we  have used the eikonal  model
to study   the role of mini-jets 
 in a QCD based framework, using as much as possible
standard QCD phenomenological inputs, and techniques.
 Our model gives a description of the data on total cross-sections in terms
of  various QCD inputs at Leading Order (LO). It is able to address two major characteristics
of the total cross-section, namely the sudden, power-like rise with energy at
the ISR and the subsequent levelling off as needed by the Froissart bound.

Our LO description  of the rise is based on  LO parton 
density functions and LO parton-parton cross-sections, implemented by soft gluon resummation of the initial state radiation, which in turn is parametrized through the same LO PDF used for the mini-jet calculation. Non-leading effects in the resummation are incorporated in a non-perturbative ansatz  for the singular, but integrable,  coupling of soft gluons in the infrared region.  Our model uses these inputs to obtain results for the total and the inelastic cross-section which are compared to the Tevatron and LHC energies. We notice that  the inelastic cross-section poses yet another challenge, as the theoretical framework in which to model this quantity is not as easily defined as is the case for the total cross-section. We  discuss this point in connection with eikonal models.  

The plan of this  paper is as follows. In Sec.~\ref{sec:ineltotal} we discuss some features of the 
 basic eikonal mini-jet model,  and discuss the limitations of the two channel eikonal model. In Sec. ~\ref{sec:model} we briefly  summarise
  the relevant details of our model, which includes soft gluon \ktresummation,  and 
   in Sect.~\ref{sec:total} we present numerical results on the   total cross-section and some of the quantities which crucially control its energy dependence.  In Sect.~\ref{sec:inel} we
apply our model to  calculate  the inelastic cross-section and  discuss the 
range of model parameters required to obtain an adequate description of  
currently available data for both the total and inelastic cross-sections
 at $7$ TeV and beyond. Our  results  are then  compared with existing data and recently measured  total and inelastic cross-sections  at LHC. 
  We end in Sect.~\ref{sec:comment} with a comment on  the limitations of   eikonal models in describing total and inelastic cross-sections.  

\section{Total and inelastic cross-sections in Eikonal mini-jet models \label{sec:ineltotal}}
Eikonal mini-jet models are an extension of Glauber's theory of diffractive 
scattering \cite{Glauber:1970jm}, with the high energy dependence driven  
by QCD processess, such as  production of low-$p_t$(mini) jets.  In addition 
to proton-proton scattering  \cite{Durand:1987yv,Durand:1988ax} and 
$p-air$ processes \cite{Durand:1988cr,Margolis:1988ws},  the eikonal minijet 
models have been applied to photon 
processes \cite{Fletcher:1991jx,Honjo:1993fi}.  QCD inspired versions 
\cite{Block:1998hu,Block:2000pg} have been used to provide a unified 
description of proton and photon processes. Our version of the eikonal 
mini-jet model uses LO  parton densities, determined from Deep Inelastic
Scattering (DIS) 
 and QCD evolution, and includes soft gluon $k_t$-resummation as a crucial 
component to access the very  large distances probed by total cross-sections. 
Our  model has been  used to analyse total cross-sections for 
$pp $ and $p\bar{p}$ \cite{Godbole:2006qk,Achilli:2007pn},
$\pi N$ and $\pi \pi$ \cite{Grau:2010ju}, $\gamma N$ \cite{Godbole:2008ex} and 
$\gamma \gamma$ \cite{Godbole:2000rh,Godbole:2008hx},  and is shown to give a 
satisfactory description of the high energy behaviour of  available data.  

Eikonal  models are formulated in the b-space representation of the scattering amplitude
\begin{equation}
F(s,t)=\int d^2 {\bf b} f(b,s)=i\int d^2 {\bf b} e^{i{\bf q}\cdot {\bf b}}
[
1-e^{
i\chi(b,s)
}
]
\label{eq:amplitude}
\end{equation}
One then gets in  impact parameter space 
\begin{equation}
\label{eq:elasticdiff}
\frac{d^2\sigma_{elastic}}{d^2{\bf b}}=|1-e^{
i\chi(b,s)
}|^2
\end{equation}
Use of  the optical theorem gives
\begin{equation}
\label{eq:total}
\sigma_{ total}=2\int d^2{\bf b}{\Re e[1-e^{i\chi(b,s)}]}=2\int d^2{\bf b}[1-\cos\Re e \chi(b,s)e^{-\Im m \chi(b,s)}]
\end{equation}
With the above equations,  and { defining  $\sigma_{inel}=\sigma_{total}-\sigma_{elastic}$} one 
   obtains
the total inelastic cross-section as
\begin{equation}
\sigma_{inel}=\int d^2 {\bf b} [1-e^{-2\Im m\chi(b,s)}]
\label{eq:inex1}
\end{equation}
The above equation presents the possibility
of  calculating the  imaginary part of the eikonal function $\chi(b,s)$, 
 by relating it to the average number of  independent 
inelastic collisions at  given values of  impact parameter b and   c.m. energy 
$\sqrt{s}$. It is in fact possible  to obtain the  above expression for the inelastic total cross-section   
through a semi-classical argument, based on the hypothesis that the scattering 
between hadrons takes place through multiple parton-parton collisions  which 
are independently distributed. This corresponds to assuming a 
Poisson distribution { around} an average number of collisions  $\nbar$, namely
\begin{equation}
P(\{n,\nbar\})=\frac{(\nbar)^ne^{-\nbar}}{n!}
\end{equation}

One then calculates the scattering at each impact parameter value $b$ between 
 the scattering hadrons and  integrates over the impact parameter space. 
Explicitly,
\begin{equation}
\sigma_{inel}(s)=\sum_{n=1}\int d^2{\bf b} \ P(\{n,\nbar\})=\int d^2{\bf b}[1-e^{-\nbar (b,s)}]
\label{eq:inex2}
\end{equation}
Comparing Eq.(~\ref{eq:inex1}) with Eq.(\ref{eq:inex2})  allows the 
identification $ {\bar n}(b,s)=2\Im m \chi(b,s)$.  Building a model for {  the 
average number of inelastic collisions then leads to  $\Im m \chi$ and}  to making 
a model for the total cross-section as well. In particular, by putting $\Re e\chi(b,s)\approx 0$ in Eq.~(\ref{eq:total}), one gets
\begin{equation}
\label{eq:1}
\sigma_{tot}(s) = 2 \int d^2{\bf b} [1 - e^{-\nbar(b,s)/2}]
\end{equation}
and
\begin{equation}
\label{eq:2}
\sigma_{el}(s) =  \int d^2 {\bf b} [1 - e^{-\nbar(b,s)/2}]^2.
\end{equation}

  Such simple mini-jet models, as described above, can be used successfully to describe the total cross-section but the parameters,  which allow to reproduce correctly the energy dependence
of the total cross-section, 
predict inelastic cross-sections which are generally  too small. Correspondingly, they predict elastic cross-sections too large and  one 
needs further input to  reproduce  
 the differential elastic cross-section. 
An  explanation, put forward in \cite{Lipari:2009rm},  is  that a simple description of the scattering in terms of just 
two channels, elastic and inelastic, or elastic and total, is not adequate and 
that (single and double)  diffractive channels need to be treated  separately. To avoid introducing more parameters,  one can insist on a two-channel model, as in   Eqs.~(\ref{eq:inex2}), (\ref{eq:1}) and (\ref{eq:2}), but in such case   the  inelastic cross-section,  calculated through these  equations in the eikonal mini-jet model,   would not include 
  correlated processes.  The reason for this comes from comparing  Eqs. ~(\ref{eq:inex1}) and  (\ref{eq:inex2}), which show that the inelastic formula, Eq.~(\ref{eq:inex1}),   
 includes only those  processes where the underlying dynamics contains no correlations between collisions.  
 
  The above observation  deserves a brief consideration.   Many models, which use the eikonal framework, apply Eqs. ~(\ref{eq:inex1}) and (\ref{eq:total}) to calculate the inelastic cross-section. The above discussion shows that, in so doing, the result will miss the contribution of correlated processes, such as diffraction. This is particularly important when extracting total cross-section data from cosmic ray experiments, which  measure    the inelastic  $pp$ cross-section \cite{Anchordoqui:2004xb}. Another case, where the inelastic cross-section matters,  is the calculation of the Survival Probability for Large Rapidity Gaps \cite{Bjorken:1992er}. Such models, \cite{Block:2001ru,Block:2011uy} or our own \cite{Godbole:2006qk}, use the probability for occurence of no inelastic collisions in $b$-space,   and caution must be exerted in interpreting what type of processes are  excluded.

 We shall return to this point at the end of the paper and investigate it in detail in future work.    In the meanwhile, we  present the results we obtain  with the model described by Eqs.~(\ref{eq:inex2}), (\ref{eq:1}) and (\ref{eq:2}) and study the role played by the  parameters in describing total 
and inelastic  cross-sections.  We shall then compare our results with the recent ATLAS Collaboration \cite{
Aad:2011eu} analysis of inelastic collisions at LHC c.m. energy 
$\sqrt{s}=7\ TeV$,  as well as with the preliminary CMS results~\cite{cmsnote}
and with the TOTEM value for the total cross-section\cite{Latino:arXiv1110.1008}.

\section{The eikonal minijet model with soft $k_t$-resummation \label{sec:model}}
The average number $\nbar (b,s)$ of collisions  can be written as the number of scattering centers per unit area $A(b,s)$ multiplied by the 
scattering cross-section $\sigma (s)$, schematically  $A(b,s)  \sigma (s)$. 
In the parton model, we need both (i) the various parton-parton cross-sections 
as well as (ii) the parton densities for an evaluation of $A(b,s)$,  
$\sigma (s)$ and hence $\nbar (b,s)$.

In minijet models, the rise of the total cross section with energy
is related to the increasing probability  of perturbative low-x parton-parton 
collisions. { Defining as    $p_{tmin}$  the threshold for perturbative treatment of parton-parton scattering}, this partonic contribution to $\sigma (s)$
 can be calculated as
\begin{equation}
\sigma^{AB}_{\rm jet} (s;\ptmin) = \int_{\ptmin}^{\rs/2} d p_t \int_{4
p_t^2/s}^1 d x_1 \int_{4 p_t^2/(x_1 s)}^1 d x_2 \sum_{i,j,k,l}
f_{i|A}(x_1,p_t^2) f_{j|B}(x_2, p_t^2)~~
 \frac { d \hat{\sigma}_{ij}^{ kl}(\hat{s})} {d p_t}.
\label{sigjet}
\end{equation}
Here, $i, \ j, \ k, \ l$ denote the partons and $x_1,x_2$ the
fractions of the parent particle momentum carried by the parton.
$\sqrt{\hat{s}} = \sqrt{x_1 x_2 s}$  and $\hat{ \sigma}$ are the
center of mass energy of the two parton system and the hard parton
scattering cross--section respectively. The numerical evaluation of this quantity strongly depends upon $p_{tmin}$ and the chosen set of Parton Density Functions (PDFs). 

The minijet model is moot regarding the region $p_t <  p_{tmin}$,
and the underlying formalism is well defined only for perturbative values
of momenta of the scattered partons for $p_t >p_{tmin}$.
  { To take into account  processes for which $p_t^{outgoing\ parton}< p_{tmin}$, one can split $\nbar$ into a soft and a hard part. i.e.  
\begin{equation}
\nbar(b,s)=\nbar_{soft}(b,s) + \nbar_{hard }(b,s)
\label{E1}
\end{equation}
 We  then  factorize each term into an overlap function $A(b,s)$ and a cross-section $\sigma(s)$, as
\begin{equation}
{\bar n}_{soft/hard}(b,s)=A_{soft/hard}(b,s)\sigma_{soft/hard}(s)
\end{equation} 
The subscript 
{\it hard} 
is meant to indicate the  
perturbative origin of the mini-jet contribution
to $\sigma_{total}(s)$ 
in calculating  $\nbar (b,s)$,  as given by Eq.~(\ref{sigjet}).  To proceed further in the
calculation of $\nbar(b,s)$, one needs to supplement this simple model 
with the knowledge of density of partons in impact parameter space as 
well as a model for $\nbar_{soft}(b,s)$.  
These issues will be   discussed below and in the next section. 
\subsection{Overlap functions in the transverse space}

In the  earliest discussions  of eikonal models~\cite{Durand:1988ax},
$A(b,s)$ was taken to be 
independent of $s$ and was obtained by taking 
the convolution of the Fourier transform of the form factors (FF) of the colliding 
particles $A,B$, given by
\begin{equation}
A^{AB}_{FF}(b)=\int {{d^2 {\bf q} }\over{(2\pi)^2}}{\cal F}^A(q){\cal F}^B(q) e^{i {\bf q}\cdot {\bf b}}
\label{eq:aff}
\end{equation}

However, these simple parametrisations of the overlap function $A(b,s)$,
augmented with a model for $n_{\rm soft} (b,s)$  failed to  
reproduce correctly the observed energy dependence~\cite{Grau:1999em}. 
{ When using in $\sigma_{jet}$ standard parton densities  determined from  Deep Inelastic Scattering (DIS),} one found an energy rise that was either too fast or started too early. An acceptable description of the data, from the beginning of the
rise to the  TeVatron energies, was not possible without 
modifying the parton densities in an {\it ad hoc} manner.

 We were able to obtain a good description of the energy behaviour   using soft-gluon $k_t$-resummation,  in a model (BN) labelled after 
 Bloch and Nordsieck \cite{Bloch:1937pw}.  It reflects the idea that any  description of the 
scattering  process between charged particles needs to include an  infinite number of soft quanta in order 
to obtain a finite result.  This  model  provides an energy dependent
form factor, with the energy dependence introduced through the kinematics of
single soft gluon emission in the $2\rightarrow 2$ parton-parton collision. In
the BN model, the parton distribution in $b$-space, is determined as the Fourier
transform  of the probability, { $d^2P(\bf K_t)$,} that the initially collinear
parton-parton pair acquire, in the collision,  an overall 
imbalance of transverse momentum ${\bf K}_t$
from  emission of an indefinite number of soft gluons. 
We proposed
\begin{equation}
A_{BN}(b)=N \int d^2P({\bf K}_t) e^{i{\bf K}_t\cdot \vecb}
\label{abnfourier}
\end{equation}
with the condition $\int d^2{\bf b}A_{BN}(b)=1$.

The soft gluon transverse momentum 
distribution can be written as
\begin{equation}
d^2P({\bf K}_t)=d^2{\bf K}_t\int \frac{d^2{\bf b}}{(2\pi)^2} exp[-i{\bf K}_t\cdot {\bf b}-h(b,q_{max})]
\label{d2pk}
\end{equation}
where $q_{max} $ is the appropriate scale for single soft gluon emission and 
the regularized function $h(b,q_{max})$ describes the overall soft gluon 
spectrum integrated over all possible values, up to  $q_{max}$, i.e.
\begin{equation}
h(b,q_{max})=\int d^3{\bar n}_g(k)[1-e^{i{\bf k}_t\cdot \bf b}]\propto \int^{q_{max}} d^2 {\bf k}_t\frac{\alpha_s(k_t)}{k_t^2}[1-e^{i{\bf k}_t\cdot {\bf b}}]\ln\frac{2q_{max}}{k_t}
\label{hdb}
\end{equation}
Eq.~(\ref{d2pk}) was proposed in \cite{PancheriSrivastava:1976tm} for soft 
emission.  
 In its  application to hard QCD processes, such as lepton pair production 
in \cite{Dokshitzer:1978yd,Parisi:1979se},  an expression for 
$h(b,q_{max})$ was proposed in which the integral of Eq.~(\ref{hdb}) extends only
down to a QCD scale $\mu$.  { This allowed} the use of the usual Asymptotic Freedom 
expression (AF) for $\alpha_s$ ($\alpha_s^{AF}$),  safely neglecting
the second term in the square bracket { in Eq.~(\ref{hdb})}. The proposal in \cite{Dokshitzer:1978yd}
has been used extensively in phenomenological analyses and can be traced back 
also to \cite{Sudakov:1954sw}, and the resulting expression for 
$ d^2P({\bf K}_t)$ is usually referred to as the Sudakov form factor. For  
processes dominated by large distances,  we have proposed that the $k_t$ 
integration be extended down to zero \cite{Nakamura:1983am}. Of course this { calls for the}  use of 
an appropriate form for the { coupling of soft gluons to the emitting quarks, when $k_t\rightarrow 0$ in Eq. ~(\ref{hdb}). This}  
strong coupling in the infrared region  must be such as { to } give rise to an integrable spectrum. We have discussed this in 
our papers, in particular in {  \cite{Corsetti:1996wg,Godbole:2004kx} and more recently in }\cite{Grau:2009qx}.

From Eqs.~(\ref{abnfourier}) and (\ref{d2pk}), our expression for $A(b,s)$,
the normalized Fourier transform of the soft gluon resummed transverse momentum distribution, is
\begin{equation}
A_{BN}(b,s)=\frac{e^{-h(b,q_{max})}}{\int d^2{\bf b} e^{-h(b ,q_{max})}}
\label{hbs}
\end{equation}
where the dependence of $A_{BN}(b,s)$ on the center of mass energy arises through   
the kinematic quantity $q_{max}$. In our simplified version of the model 
proposed in \cite{Corsetti:1996wg}, this is obtained by averaging over the 
parton densities.
\comment{
The details of the function $h(b,s )$  are discussed in a section dedicated to 
soft $ k_t$-resummation { where is this section going to be placed?}}
Thus, the rising -high  energy part- of the average number of collisions, is 
written  as
\begin{equation}
\nbar_{hard}(b,s)=A_{BN}(b,s)\sigma_{jet}(s,p_{tmin})
\end{equation}
\subsection{Scales and parameters}
While we refrain from giving many details of the model, as they have been 
discussed carefully in our different 
publications~\cite{Grau:1999em,Godbole:2004kx,Grau:2010ju}, it is necessary
that we  give  a brief discussion of the crucial parameters and the different
scales involved, so that the physics issues become clear. Along 
with  $p_t$, the transverse momentum 
of the partons involved in the basic $2 \rightarrow 2$ scattering, another
transverse momentum  { variable} of relevance is $k_t$, the transverse momentum
of the soft gluons emitted during the hard scattering process. This somewhat 
artificial separation, allows a clear delineation of different theoretical 
issues involved. The role of $\ptmin$ in the calculation of 
the hard part of the partonic cross-sections has been already  explained.
{ For the variable $k_t$, a}n important role is played by 
 $q_{max}$, 
the maximum  value allowed  
 to  single soft gluon emission during the collision 
between  two partons.{ Using the kinematics of single gluon emission, this quantity can be written as}
\begin{equation}
q_{max}(\shat,y,Q^2)=\frac{\sqrt{\shat}}{2}
(1-\frac{Q^2}{\shat})
\frac{1}{\sqrt{1+z\sinh^2 y}}.
\label{eq:qmax}
\end{equation}
Here, $ Q^2$ is the squared  invariant mass of the 
outgoing  parton pair, each parton  with transverse momentum $p_t$.
In the no-recoil approximation, $z=Q^2/\shat$ and $y$ is the rapidity 
of the outgoing partons \cite{Chiappetta:1981bw}. This scale, which 
affects the final result mostly through logarithms, is a semi-hard 
scale   akin to $\ptmin$  and it is also of the same order of magnitude.

Depending 
on the energy scale of the process, as discussed at length
in \cite{Grau:1999em}, the calculation of the emission of soft gluons of  
momentum $k_t\le q_{max}$ requires a special treatment. In our model, in order to be able to describe very large distance contributions to the total cross-sections, we have retained the second term in the square bracket in Eq.~(\ref{hdb}). Thus, yet
 another scale, in the problem at hand, is the impact 
parameter $b$ and special care is needed in discussing the effects in the
infrared region when  $k_t b \sim 1$. When $b$ becomes very large and  $k_t\le 1/b< \Lambda_{QCD}$, 
the calculation of $h(b,q_{max})$, and hence the overlap 
function, { requires an ansatz for}   the behaviour of $\alpha_s$ in the far infrared where one
can not use the usual expression at large scales. 
  In the infrared region, our ansatz is to use
\begin{equation}
\alpha_{eff}=\alpha_s^{IR} =c \left( {\Lambda^2_{QCD}} \over {k_t^2} \right)^p
\label{alphair}
\end{equation}
namely an expression which is singular but integrable, provided $p<1$.  We stress here that to access very large distances, we consider  the very soft gluons for which  $k_t \rightarrow 0$, which in turns implies resummation and, in the continuum limit, integration. Thus what matters is not the limit of the coupling for one soft gluon, but the integrability, a concept also put out by  Dokshitzer
 \cite{Dokshitzer:1998nz} and  used in jet analysis \cite{Dokshitzer:1999sh}.

 As discussed below,
   the expression  in Eq.~(\ref{alphair}) mimics the confinement 
dynamics.  To perform the full 
integral of Eq.~(\ref{hdb}), which spans from the infrared to the beginning of the 
AF region,  we  then use an interpolating expression {  for $\alpha_s$, which reduces to the correct AF limit at 
large scales and   to Eq.~(\ref{alphair}) in the infrared region  \cite{Corsetti:1996wg}. 

The parameter $p$ captures the physics at  scales $k_t \lesssim \Lambda_{QCD}$ and 
affects the high energy behaviour of the total cross-section in a complex 
manner. As shown in ~\cite{Grau:2009qx}, its value  has to be between $1/2$ and $1$. 
As a consequence, the very large b-limit of  the function $A_{BN}(b,s)$, i.e.
\begin{equation}
A_{BN}(b,s)\sim e^{-h(b,s)}\sim e^{-(\bar {\Lambda} b)^{2p}}
\end{equation}
shows   the impact parameter distribution falling at its fastest    as  a gaussian ($p=1$) and at its slowest
 as  an exponential ($p=1/2$). The scale $\bar \Lambda\propto \Lambda_{QCD}$  includes a mild energy dependence   through the scale $q_{max}$, as well as a residual dependence upon the parameter $p$.   This behavior in impact parameter space joined with the high energy behaviour of the mini-jet cross-sections, $\sim s^\varepsilon$,    was shown to lead to  an  asymptotic behaviour of the total cross-section consistent with the Froissart bound, namely
\begin{equation}
\sigma_{T}\approx \frac {2\pi } {{\bar \Lambda}^2} [\varepsilon \ln \frac {s} {s_0}]^{1/p}
\label{froissart1}\end{equation}
upto leading terms in ($\ln s$) \cite{Grau:2009qx}.

\subsubsection{About the singularity parameter $p$}
Let us briefly discuss the physics implications of the choice of a 
particular value of the singularity parameter $p$.  In our model, $p$ 
describes how singular 
$\alpha_{eff}$ is in the IR region and can be related to confinement dynamics. This is seen through the spatial-potential 
obtained through the Fourier transform of the one-gluon exchange potential 
generated by our effective coupling, namely
\begin{equation}
V(r) \sim r^{2p-1}\ \ \ \ \ \ r\rightarrow \infty
\end{equation}
For  $p=1/2$  the dressed one-gluon exchange potential is (essentially) a constant , whereas 
for $p=1$ it is linearly rising. 
In ~\cite{Grau:2009qx}, we 
showed that  the parameter $p$ has to be between $1/2$ and $1$, so that the
corresponding potential is confining and  $\alpha_{eff}$  at small scales is 
singular but {\it integrable}. 
 On the other hand,  given our ignorance of the actual confinement dynamics,  we shall use it as a free parameter which  can interpolate between a fully confining potential, linearly rising when $p=1$, and the inelastic scenario,  in which
 partons are free to 
hadronize,   with $p=0.5$ and the potential constant at infinity. 

 What we have given above  
  is 
certainly  too simple an argument, but it is not in 
contradiction with our understanding of the role played by the infrared 
singularity.  From our phenomenology of the total cross-section \cite{Godbole:2006qk,Achilli:2007pn,Grau:2010ju, Godbole:2008ex}, we have found that the value    $p\sim 3/4$ gives acceptable descriptions of high energy data. Such a value would then  be consistent with  
a rising potential.

\subsubsection{ About the PDFs}
  We consider the PDFs in our model to be phenomenological QCD  parameters  through which we describe the two high energy effects which are input to the eikonal: the rise at the beginning and the slowing down towards  asymptotia. So far, ours is a Leading Order (LO) description of these two effects. In our parametrization of these effects  we use LO densities able to go down to very small x-values and $Q^2\sim 1 \ GeV^2$ (see below). In our first application of the model in \cite{Grau:1999em}, we have found that GRV densities \cite{Gluck:1991ng}, when combined with our resummation expressions,  provide a parametrization of the  mini-jet effects  which leads to  a  good description of total cross-section data. Through the years, other LO PDFs and many NLO and NNLO pametrizations have appeared. To use some of them in our model would be inconsistent and imply major modifications, as  we explain below. Yet another possibility will be to employ the
LO* distributions~\cite{Sherstnev:2007nd}, suggested  for use
with the LO Monte Carlos. In future work, this and other possibiilties will be examined. Here we present the   reasonably good LO description
of the total cross-section, obtained so far in our model  in terms of the
various inputs such as LO  PDFs, $p_{tmin}$ and $p$, and show how far it can take us in probing the
long distance behaviour of QCD.

 To be of significance in QCD descriptions of scattering processes, mini-jet 
models should   use Parton Density Functions determined from Deep Inelastic 
Scattering. There are many PDFs, and not all of the available sets can be used 
in a  model  like ours, in which the rise of the total cross-section is due
only to mini-jets.  To have the rise appearing already  at ISR energies, one needs  $Q_0^2=p_{tmin}^2\simeq 1 \ GeV^2$ and the PDFs we use  must be valid down to such a   low $Q_0^2$ value. Notice that  we can only  use LO  PDFs,   because in our model, the bulk of NLO contribution comes from soft gluon resummation, and using NLO versions might produce a double counting.  The chosen PDFs must also be able to describe very low values of $x$, the energy fraction carried by the incoming partons. In particular most important are the low-$x$ gluon densities. Notice however that as the energy increases and  lower and lower  $x$-values  are accessed, soft gluon resummation softens and tames the rise of low-$x$ gluon contribution. This contribution is not included in the low-$x$ phenomenology of current PDFs and this may pose a problem since the low $x$-behaviour of existing densities is modified by our model resummation inputs. The claim for instance that the low $x$-behaviour of some of the densities such as  GRV is wrong, needs to be seen in light of resummed contributions like the ones we are considering :  soft gluon effects are a NLO effect, but  resummation to all orders down into the infrared is not included in the low-x behaviour of  PDFs, neither in the LO nor in the NNLO parametrizations. Since such effects, in our opinion, are crucial for inclusion of QCD in descriptions of the total cross-section, we have resorted to use only LO parametrizations for all QCD effects. Even resummation is treated to LO order, since the argument of the integral in Eq.~(\ref{hdb}) does not include NLO terms  and $q_{max}$ is also averaged only over leading valence quak effects (see below). On the other hand, as discussed in the Introduction, the bulk of non-leading effects in the resummation is incorporated in our  non-perturbative ansatz  for the singular, but integrable,  coupling of soft gluons in the infrared region, Eq.~(\ref{alphair}).

Our model uses the same approach also in dealing with the calculation of the quantity $q_{max}$. Using the values   $y=0$, $p_t=p_{tmin}$, we   average the quantity given in Eq.~(\ref{eq:qmax})  over the parton densities, i.e. we use 

\begin{equation}
<q_{max}(s)>={{\sqrt{s}} 
\over{2}}{{ \sum_{i,j}\int {{dx_1}\over{ x_1}}
f_{i/a}(x_1)\int {{dx_2}\over{x_2}}f_{j/b}(x_2)\sqrt{x_1x_2}
\int_{z_{min}}^1
 dz (1 - z)}
\over{\sum_{i,j}\int {dx_1\over x_1}
f_{i/a}(x_1)\int {{dx_2}\over{x_2}}f_{j/b}(x_2) \int_{z_{min}}^1 (dz)}}
\end{equation}
with $z_{min}=4p^2_{ptmin}/{\hat s}$.} 
 As discussed in \cite{Godbole:2004kx}, we only include the leading soft gluon resummation from valence quarks and not from  low-$x$ gluons.  
 The emission of soft radiation only from valence quarks, and thus the  determination of $q_{max}$,  is also part of our LO parametrization, and  follows the idea that in resummation in QED, the leading terms correspond to the external legs. Although  in the impulse approximation gluons are treated as free particles,  gluons in gluon-gluon collisions  must partake   of the initial acollinearity imparted by the valence quarks. It would be inconsistent to imagine that  the acollinearity due to initial state radiation  from valence quarks does not reflect on all subprocesses.  This is why at LO we consider  this to be  the major effect. The introduction of emission from interacting gluons will have to be considered but  the  formalism is more complicated, and we first want to see how far the LO model can take us.

 Among the available PDFs satisfying these requirements, there are  GRV,  MRST and CTEQ densities. The latter, in our model, do not yield   a rising cross-section past the TeVatron data.  Within  our approach, the reason lies in the fact that the quantity $q_{max}$, calculated with CTEQ,   rises with energy so much that  the taming effect from soft gluons, which  reflects the initial   acollinearity of the partons,  does overcome the rise of the mini-jet cross-sections. We have shown the details of this exercise in \cite{Pancheri:2007rv}  and work is in progress towards a better understanding of how to use CTEQ in our model. 
  To summarize,  our choice of PDFs is a {\it parametrization} of LO QCD effects within an eikonal mini-jet model with soft gluon resummation in the infrared region:  as such, densities like CTEQ do not describe the total cross-sections and cannot be used in the model in its present formulation. On the other hand, both GRV as well as MRST LO densities are well suited to describe all the data up to the Tevatron and including cosmic rays, and can thus be used for further investigation.

 NLO corrections to our model would most certainly change the choice of model parameters, of course. An  NLO version of our model might      clarify some of the issues and  a full investigation
 in this direction is in progress.

\vspace{1cm}

 Thus, at   the end, the relevant parameters 
for evaluation of $\bar n_{hard} (b,s)$ are  $\ptmin, p$ and the chosen 
PDF set. We call them the  high energy (HE) parameters. 
 { As noted,} 
 the scale $q_{max}$, the upper limit in the soft $k_t$ integration, 
is not an independent parameter.  
Its value is determined by the kinematic variables defining the parton-parton
sub-process, namely  $p_t$, the rapidity $y$ and $\shat$.
In our model, for the calculation of $q_{max}$, we choose $y=0$, $p_t=p_{tmin}$. The scale $q_{max}$, used 
in obtaining the impact parameter distribution $A_{BN}(b,s)$,  is  then 
calculated as an average over the parton densities in the colliding hadrons. 
 Once the parameters 
 $p_{tmin}$ and the PDF set  are chosen,  $\sigma_{jet}$ and $q_{max}$ can be calculated.  One then chooses a value for the parameter $p$, and using   $q_{max}$, one can completely determine $A_{BN}(b,s)$.  The calculation of $\bar n_{hard} (b,s)$ then follows from choosing $p$, $p_{tmin}$ and the PDFs \cite{Corsetti:1996wg,Grau:1999em}.} 
\section{The total cross-section \label{sec:total}}
{ The mini-jet model we have outlined in the previous section describes only  the rising part of the total cross-section. While the hard term $\bar n_{hard}(b,s)$ dominates at high energy and is 
supposed to include the dynamical mechanism of the rise of the cross-section 
with energy,  a substantial part of the total cross-section is also due to  
soft processes with $p_t\le p_{tmin}$,  according to our model. Due to our limited understanding of the QCD dynamics of this part of the cross-section, $\nbar_{soft}(b,s)$ is parametrized. We start with factorization
\begin{equation}
\nbar_{soft}(b,s)=A_{soft}(b,s)\sigma_{soft}(s)
\end{equation}
and  have 
used two models to calculate $\nbar_{soft}$. The two models use two different  expressions  for the overlap functions and 
 different ans\"atze for the energy 
dependence of the soft cross-sections.  
 
 The details of these two models are discussed in Appendix \ref{sec:appendix}, where we also give values  of the parameter sets we use. We stress here the important fact that our parametrizations for ${\bar n}_{soft}(b,s)$ do not include any term rising with energy, but only constants and decreasing terms. Nowhere we impose in the model any logarithmic growth. The rise is always produced by the mini-jets.

Notice that  Eq.~(\ref{E1}) 
can be written as
\begin{equation}
\nbar(b,s)={A_{mean}(b,s)}
[{\sigma_{soft}(s) +\sigma_{jet}(s)}]
\end{equation}
with the overall mean impact parameter distribution defined as
\begin{equation}
A_{mean}(b,s)=\frac{A_{soft}(b,s)\sigma_{soft}(s) +A_{BN}(b,s)\sigma_{jet}(s)}{\sigma_{soft}(s) +\sigma_{jet}(s)}
\end{equation}
thus allowing for comparison with other models.
In { the four plots  of} Fig.~\ref{fig:bfigs} 
\begin{figure}[!]
\resizebox{0.4\textwidth}{!}{
\includegraphics{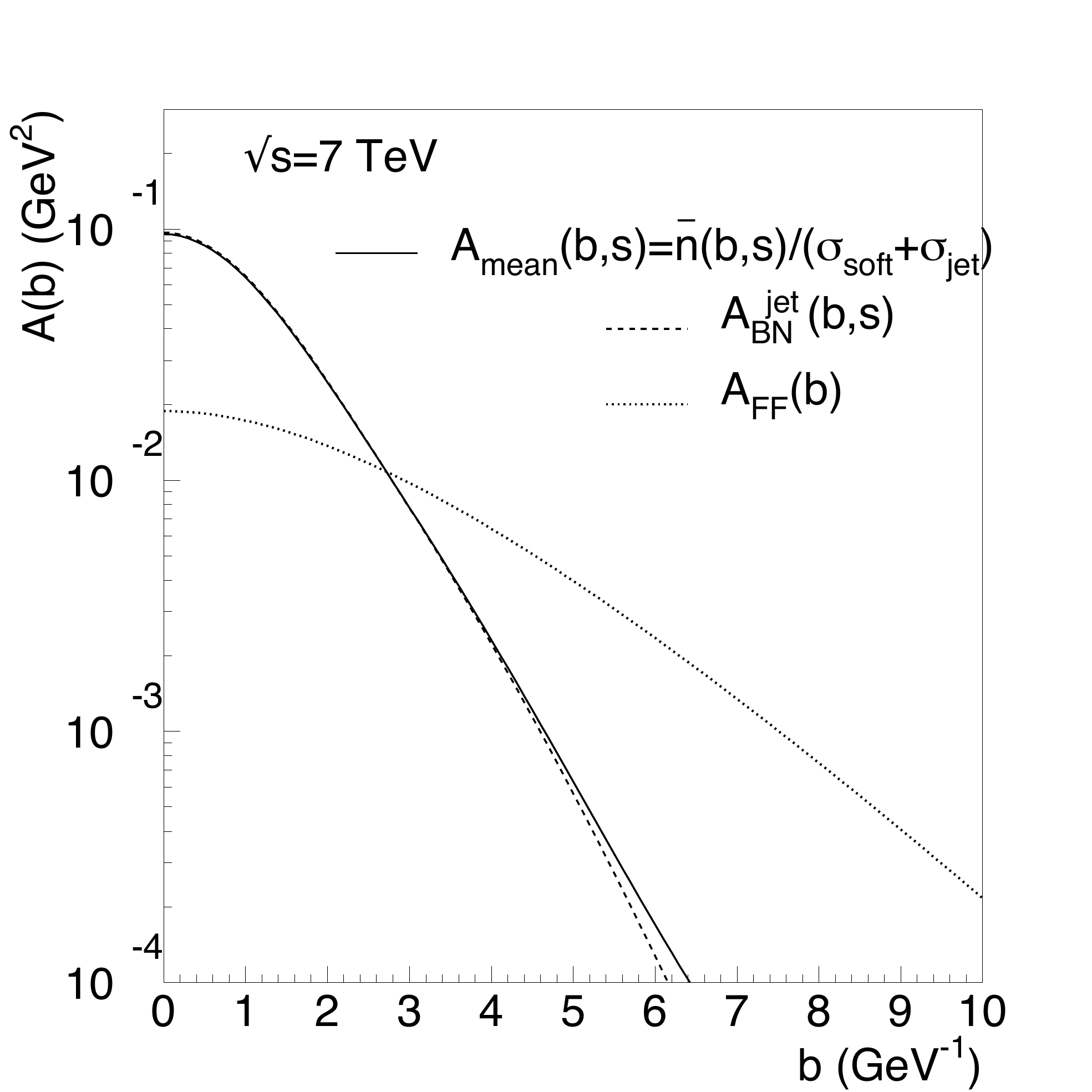}}
\resizebox{0.4\textwidth}{!}{\includegraphics{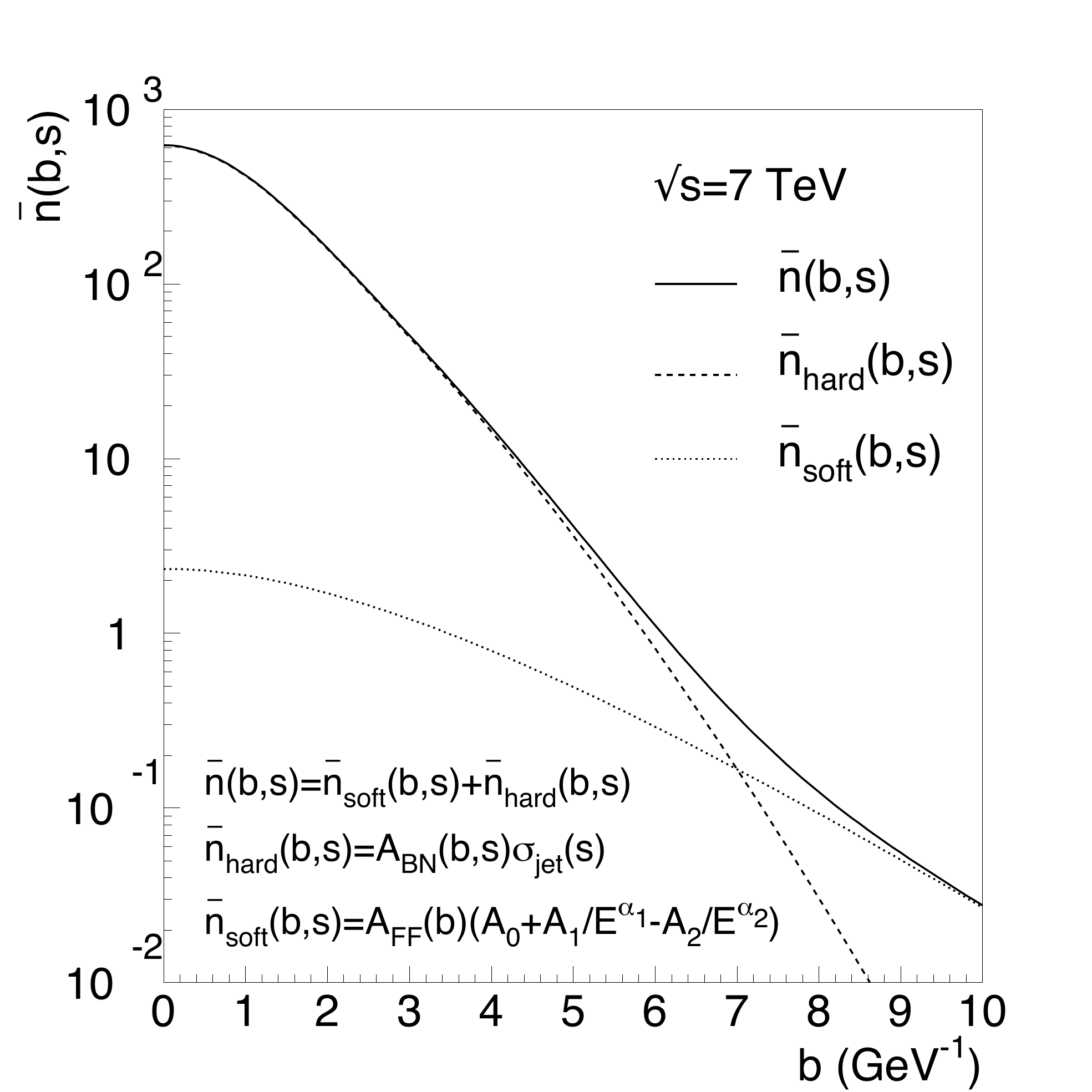}}
\resizebox{0.4\textwidth}{!}
{\includegraphics{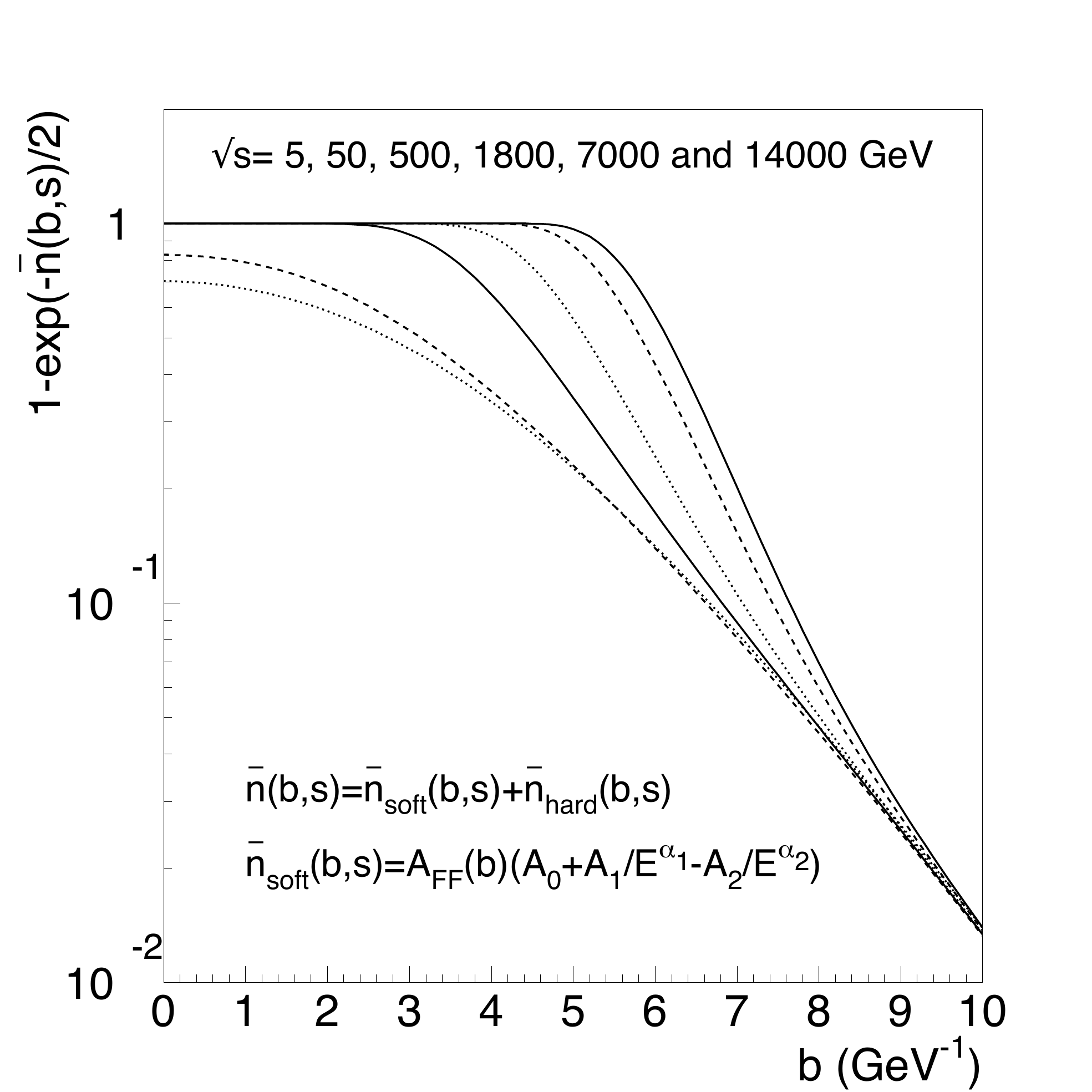}}
\resizebox{0.4\textwidth}{!}
{\includegraphics{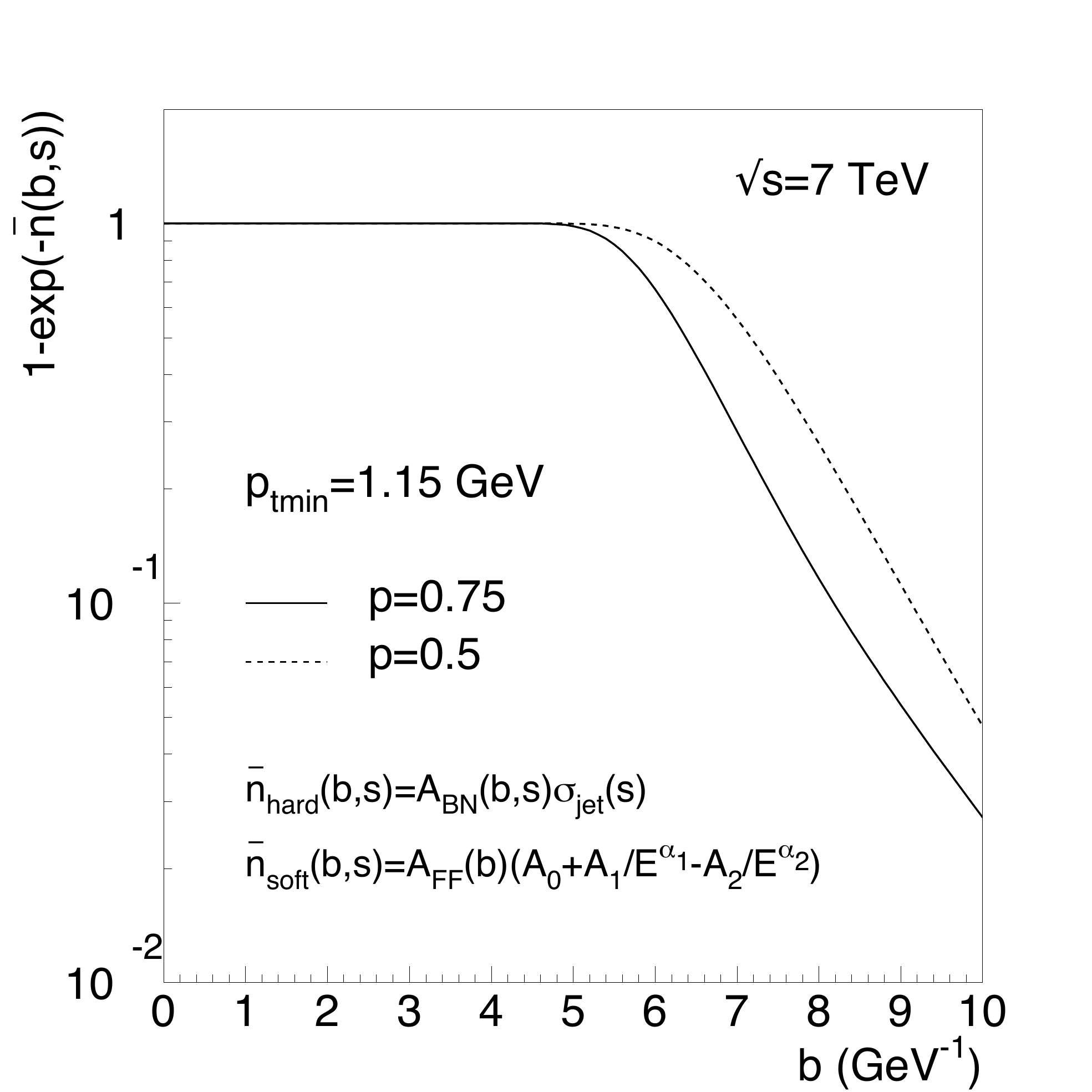}}
\caption{\small The impact parameter dependence of  different  quantities relevant for total cross-section calculations are shown.
Top left : the normalized Fourier transform of the soft gluon
$k_t$-distribution, which gives the overlap function in proton-proton
collisions, averaged over densities, as well as over  hard and
soft processes. Top right: the average number of collisions as a function
of the impact parameter. Bottom:   
at left, 
 elastic scattering amplitude  in  $b$ space, in the approximation $\Re e\chi \approx 0$ at various c.m. energies, and, at right,    the probability for inelastic collisions  at $\sqrt{s}=7\ TeV$, namely 
the integrand for   the inelastic cross-section. The high
energy parameters
 are $\ptmin = 1.15$ GeV,  GRV densities \cite{Gluck:1991ng} and $p=0.75$, except for the dashed curve in the bottom  right panel, where 
$p = 0.5$.}
\label{fig:bfigs}
\end{figure}
we show the 
model prediction for { this function and other}  impact parameter distributions of 
relevance to the total and inelastic cross-section calculation.  In all four figures, we used Model II for $\nbar_{soft}(b,s)$. Using Model I, gives very similar results.

In the two upper plots of Fig.~\ref{fig:bfigs} we show   the behaviour of $A_{mean}(b,s)$  and of the average number of collisions, ${\bar n}(b ,s)$ for  
proton processes as a function of $b$ at $\sqrt{s}=7 \ TeV$. Comparison is shown between the overall distributions and the 
contribution from soft and hard components of the imaginary part of the 
eikonal function. The bottom plots
show the  elastic amplitude in the approximation $\Re e \chi(b,s)\simeq 0$,  and }the integrand  for the 
inelastic cross-section calculation  in the 
two dimensional $b$ space. These plots  (bottom)  clearly show  
the cut-off behaviour discussed in ~\cite{Grau:2010ju}, resulting from  our soft $k_t$-resummation model in the infrared region.  It is interesting to compare, in the bottom left  figure, the behaviour at low c.m. energies with the high energy LHC  behaviour, which is more and more approaching a Fermi function in $b$-space.This cut-off behaviour
is what  transforms  the power-like behaviour of the mini-jet cross-sections 
into a logarithmic energy dependence coherent with the limits imposed by the 
Froissart bound \cite{Grau:2009qx}. 

The  bottom figure at right  illustrates the role played by the parameter p in our model. We notice that  the plateau extends  to include  larger $b-$values as $p$ decreases  from the value $p=0.75$ to $p=0.5$,  indicating a larger interaction region when $p$ is smaller. i.e. when the infrared singularity, accessed by soft gluon resummation, lessens. 
 In other words, as $p$ decreases from $0.75$ to $0.5$, the less confining behaviour for soft gluons in the infrared   enlarges  the interaction region, whose extension is roughly determined by the value where the plateau drops. In Sect.~\ref{sec:inel},  we shall compare the data for $\sigma_{inel}\equiv \sigma_{total}-\sigma_{elastic}$ with the inelastic cross-section from the model,  Eq.~(\ref{eq:inex1}), and we shall find that  the value $p\sim 0.75$, which describes total cross-section data,  falls short  of the totality of inelastic  data for energies up to the TeVatron.  We shall then resort to  
 a smaller $p$ value to describe them.

We close this discussion of the distributions  { in } impact parameter space, with a few general remarks. As is evident from the bottom
curves in  Fig.~\ref{fig:bfigs}, at
high energy the integrand in impact parameter
space is a constant until a certain value $b_{max}(s)$ (approximately
about { one} Fermi at $\sqrt{s}\ =7\ TeV$),  and then there is a tail from
the peripheral collisions. As the initial energy of the two protons
increases, the plateau is extended and the tail becomes sharper. In
the extreme high energy  limit, the distribution is expected to
become sharp (a perfect Fermi function) and in this perfect hard disk
limit the elastic cross-section should become one-half of the total
cross-section. While the trend of the data so far is towards an
increase in this ratio, the approach is extremely slow and not likely
to be reached in the near future.

\begin{figure}
\centering
\vspace{-2cm}
\resizebox{1\textwidth}{!}{
\includegraphics{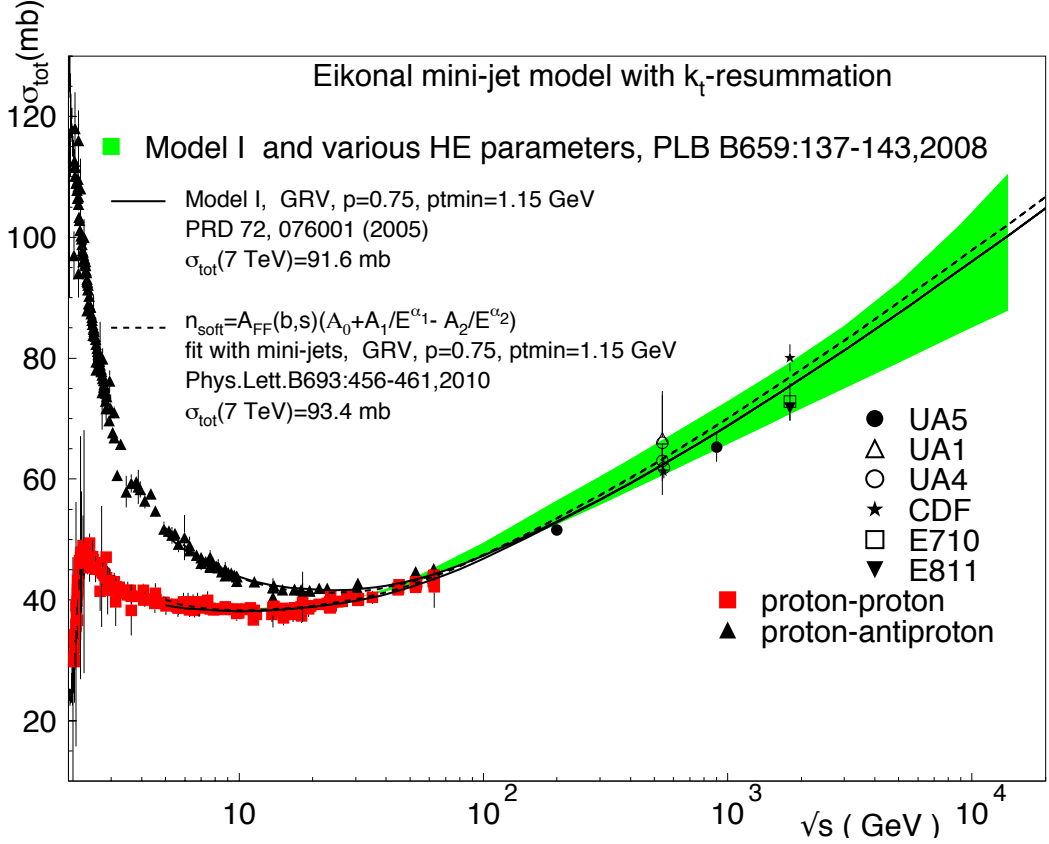}}
\vspace{-5cm}
\caption{Total \pp \  and \pbarp \ cross-section calculated with our model and 
compared with data. Values for $\sigma_{total}$ at 7 TeV at LHC are indicated.}
\label{fig:total7tev}
\end{figure}

We are now in a position to exhibit our results for the total cross-section. In 
Fig.~\ref{fig:total7tev} we show two black lines and a band: the full black  and the  { green} band are  obtained with Model I for $\bar n_{soft}$  { and have been already presented \cite{Godbole:2004kx,Achilli:2007pn}.}  
The dashed black line  was obtained  with Model II, and the low energy fit includes the mini-jet contribution. For all the fits, we have used  LO PDFs which  allow $Q^2$ values as low as 
can be reached for the chosen  $p_{tmin}^2$ as well as the corresponding low
$x$ values that are reached. 
 As discussed earlier, we use here
 the following 
parameterisations: GRV\cite{Gluck:1991ng},  GRV94 \cite{Gluck:1994uf}, 
GRV98 \cite{Gluck:1998xa}, MRST
\cite{Martin:1998sq}.
The two curves, Model-I and Model-II,  use the same set of high energy parameters, GRV densities for the  mini-jets, $p_{tmin}=1.15\ GeV$ and $p=0.75$ for the infrared singularity 
parameters. 
The
 green  band is obtained varying slightly the high energy parameters, by no 
more than a few percent,   and  using different partonic densities.  
At present LHC energies, $\sqrt{s}=7\ TeV$, we obtain 
$\sigma_{total}=91.6  \ + 6.5\ -9.7 (mb)$ with  Model I .
The results for   this model at the two LHC energies $ \sqrt{s}=7,\ 14\ TeV$, are summarized in Table~\ref{tab:andrea}. For 
Model II, we have only calculated the value with GRV 
densities, $p_{tmin}=1.15\ GeV$,    and $p=3/4$ , and the value
$\sigma_{total}(\sqrt{s}=7\ TeV)=93.5\ mb$, lies within the indicated band of 
 Model I. Our actual ignorance of a fundamental QCD description of 
the low energy total cross-section dynamics makes further studies of the 
difference between the models, quite irrelevant.  However  the result with low energy fits inclusive of mini-jet contribution, i.e. model II, should be preferred.

\begin{table}
\begin{footnotesize}
\begin{center}
\begin{tabular}{|c|c|c||c|c|c|c|}
\hline
&&&&&\\
PDF&$p_{tmin}$ (GeV) &$\sigma_0(mb)$& p&$ \sigma_{tot}$&$\sigma_{tot}$ (mb)\\
        &                                  &                        &   &$\sqrt{s}=7\ TeV$& $\sqrt{s}=14 \ TeV$\\
\hline
GRV    &1.15&48&0.75&91.6&100.3\\
\hline
GRV94&1.10&46&0.72&95.9 &103.9\\
             &1.10&51&0.78&83.3 &89.7\\
\hline
GRV98&1.10&45&0.70&94.9& 102.0\\
             &1.10&50&0.77&82.1& 87.8\\
\hline
MRST(72)& 1.25&47.5&0.74&85.9 &95.9\\
                   & 1.25&44&0.66&97.9& 110.5\\
\hline
\end{tabular}
\caption{
Values of \protect{$\sigma_{tot}$} for
 $p_{tmin}$, $\sigma_0$ 
 and $p$ corresponding to different
parton densities in the proton, for  which our  model \cite{Achilli:2007pn}
 gives a satisfactory description of the total cross-section. The values are obtained with Model I (see Appendix) for $\nsoft$, the average number of  soft collisions.}
\label{tab:andrea}
\end{center}
\end{footnotesize}
\end{table}

 Having thus chosen the set of HE parameters which adequately describe the existing total cross-section data, we now proceed to  investigate  how   the eikonal function, determined by these parameters, describes data for the inelastic cross-section.

\section{The inelastic proton-proton  cross-section \label{sec:inel}}
In this section we  present the results of our model for the inelastic cross-section. 
 It is worth noting that a
nontrivial  bound on the inelastic cross-section, based on general analyticity
arguments has been obtained only very recently~\cite{Martin:2009pt}, 
unlike its counterpart for the total cross-section which has been
known for quite some time \cite{Froissart:1961ux,Martin:1962rt,Lukaszuk:1967zz}.

The inelastic 
proton-proton cross-section is of particular interest in cosmic ray physics, 
as it determines the total $p-air$ cross-section \cite{Anchordoqui:2004xb}. 
The knowledge of this inelastic cross-section  at very high energies
 enters the simulation of   cosmic ray air showers 
  and is used for extracting  the total $pp$ cross-section from cosmic ray data.
  However, unlike the total and the elastic case, the  inelastic cross-section is not uniquely defined and data for it suffer from usage of different cuts imposed in the analysis which lead to inlcusion of differing amounts of diffractive
contribution to the measurement. In principle, the least ambiguous way to define data for the inelastic total cross-section is  through the definition
\be
\sigma_{inelastic}^{exp}\equiv \sigma_{total}^{exp}-\sigma_{elastic}^{exp}
\label{eq:sigmaineldat}
\ee
Models can then be compared with data in the available range of energies, starting from $\sqrt{s}\simeq 1-2\ GeV$ up to the TeVatron data, and now to the LHC data at $\sqrt{s}=7\ TeV$.

\begin{figure}[h]
\centering
\vspace{-2cm}
\resizebox{1\textwidth}{!}{
\includegraphics{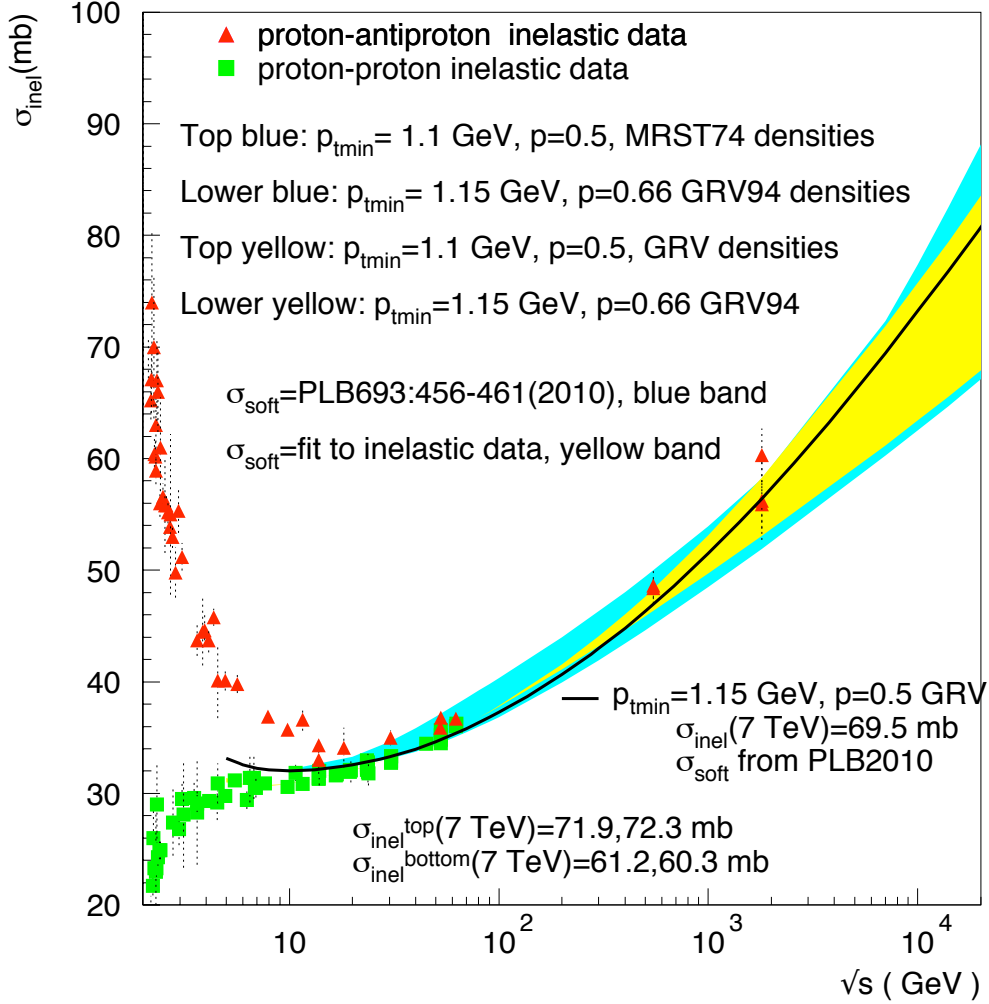}}
\vspace{-2cm}
\caption{Comparison between data for $\sigma_{total}-\sigma_{elastic}$ with results of our model. Values for the parameters used for the band and the curve are indicated.}
\label{fig:inelbands}
\end{figure}

 As a first step towards analysing the inelastic
cross-sections,  one  then needs to extract the values for 
$pp$ scattering, as the difference between the total 
and the elastic cross-sections. The error of this procedure is obtained by 
combining the  errors in quadrature.   Once the data points are obtained 
in this fashion, one can try to confront our model predictions with  them. 
Note that there are no data for $pp$ scattering beyond the ISR energies \cite{Pdg}. 
However, the increasing dominance of gluon-gluon scattering  with the rising  
energy should allow us to use the \pbarp \ data from CERN $S{\bar p}pS$ \cite{Bozzo84} and 
from the TeVatron \cite{Amos90,Amos92,Abe93,Abe94,Avila99}, as a guidance in our analysis. 
{ We notice that in \cite{Amos90}, in addition to the total and elastic data, the collaboration also presents a value for the inelastic cross-section at $\sqrt{s}=1800\ GeV$. There are therefore two possible values to fit, $\sigma_{total}-\sigma_{elastic}=56.2 \pm 3.5\ mb $ (taking
$\sigma_{tot}$ from \cite{Amos92} and $\sigma_{el}$ from \cite{Amos90}) or $\sigma_{inelastic}=55.5 \pm 2.2 \ mb$ which is the direct measure
 presented in \cite{Amos90}. The two values agree within the errors. In our estimate,  for consistency with all the other data points, which were obtained  in the same way, { we use the value  obtained by subtraction.  For the same reason, i.e.   consistency with the subtraction procedure we have outlined},  we do not use results from \cite{Augier:94}. Thus, in all the figures to follow, the term {\it inelastic data} corresponds to data obtained from the difference between total and elastic cross-section, as per Eq.~(\ref{eq:sigmaineldat}) and can be directly compared with Eq.~(\ref{eq:inex1}).

As was the case 
for the total cross-section, before studying the 
high energy behaviour, the soft part of the eikonal needs to be chosen. 
We shall use the Form Factor model, Model II, with two different choices
for the soft cross-section $\sigma_{soft}$.  In the first,  Model II-A, the soft part of the eikonal, namely ${\bar n}_{soft}=A_{FF}\sigma_{soft}$,
is the same as the one entering the total cross-section estimates shown by the  dashed line in Fig~\ref{fig:total7tev}, while in the second, Model II-O, we use an independent fit to the low energy inelastic  $pp$ cross-section data with $\sqrt{s}>5$ GeV, which also includes  the mini-jet contribution. As in the case of the total cross-section, 
at high energy, the cross-section  behaviour is rather insensitive to this choice and is mainly 
controlled by the PDF's and the other high energy parameters $p, \ptmin$.

  As anticipated  in Sec.~\ref{sec:ineltotal},
 we find that the high energy data,  
   cannot   be described by   the same set of HE parameters  determined from  Eq. ~(\ref{eq:1}) for the total cross-section,  namely GRV and MRST PDFs, $p_{tmin}$ and $p$.  
     In particular,  in Fig. ~\ref{fig:september24-tot-inel-GRV} we show a comparison between the inelastic data and the results of our model for the set of HE parameters GRV, $p_{tmin}=1.15\  GeV, p=0.75.$
   \begin{figure}
   \vspace{-1cm}
  \resizebox{0.9\textwidth}{!}
  { \includegraphics{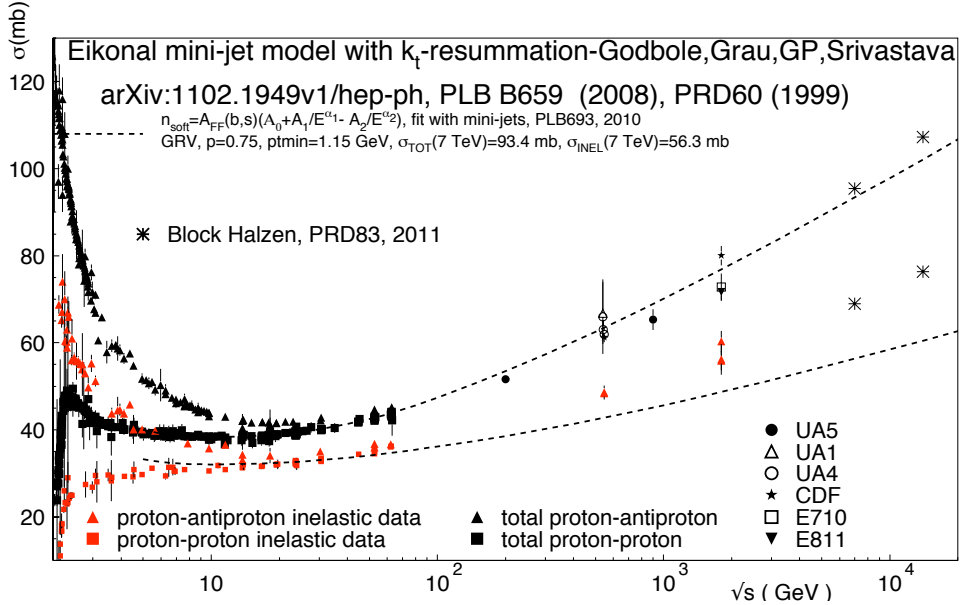}}
  \vspace{-6cm}
  \caption{The total and inelastic cross-section as described by our model when using the same set of high energy parameters in the eikonal. Results are compared with pre-LHC data and an estimate from \cite{Block:2011uy}.}
  \label{fig:september24-tot-inel-GRV}
  \end{figure}
    More generally,  varying $p_{tmin}$ and the densities as in Table ~\ref{tab:andrea} within the same ranges as we  did for the total cross-section, we find that the $p$-values in Table ~\ref{tab:andrea} always give results short of the pre-LHC data. This is consistent with the observation in Sec.~\ref{sec:ineltotal}, namely that the eikonal mini-jet model with two components  will only describe uncorrelated processes, while the {\it inelastic data} from Eq.~(\ref{eq:sigmaineldat}) include also collisions in diffractive and other similarly correlated regions.
   On the other hand,  everything else being equal, a reasonable description is obtained by letting  the parameter $p$  vary between the minimum  value consistent with the model \cite{Grau:2009qx}, $p=0.5$, and $0.66$,  the highest value which can possibly be consistent with
 the total cross-section band in \cite{Achilli:2007pn}. We show this analysis in Fig.~\ref{fig:inelbandagneseolga}.
\begin{figure}
\centering
\vspace{-1cm}
\hspace{-1cm}
\begin{minipage}{0.5\textwidth}
\includegraphics[width=1\textwidth]{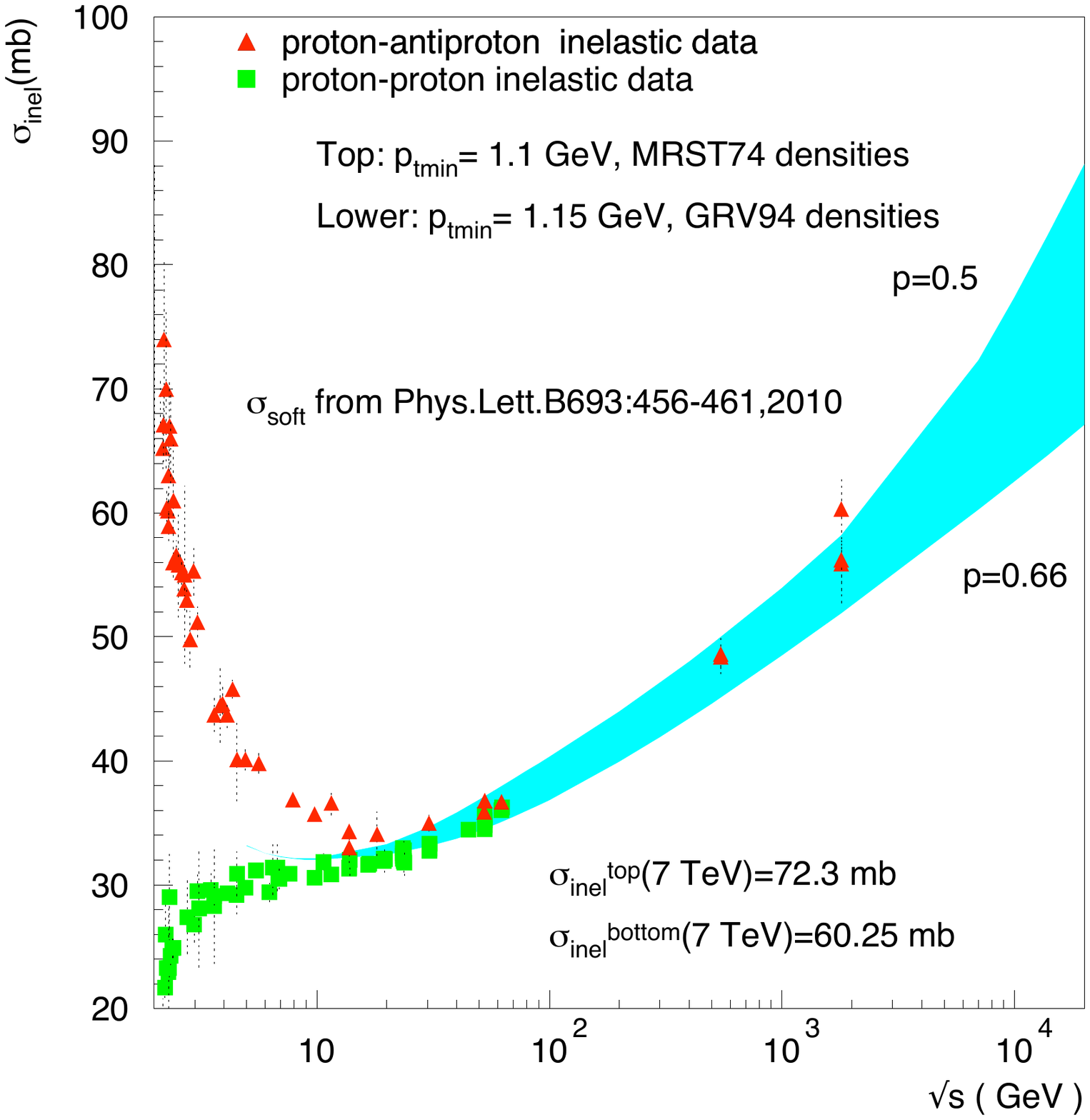}
\end{minipage}
\hspace{-1cm}
\begin{minipage}{0.5\textwidth}
\includegraphics[width=1.\textwidth]{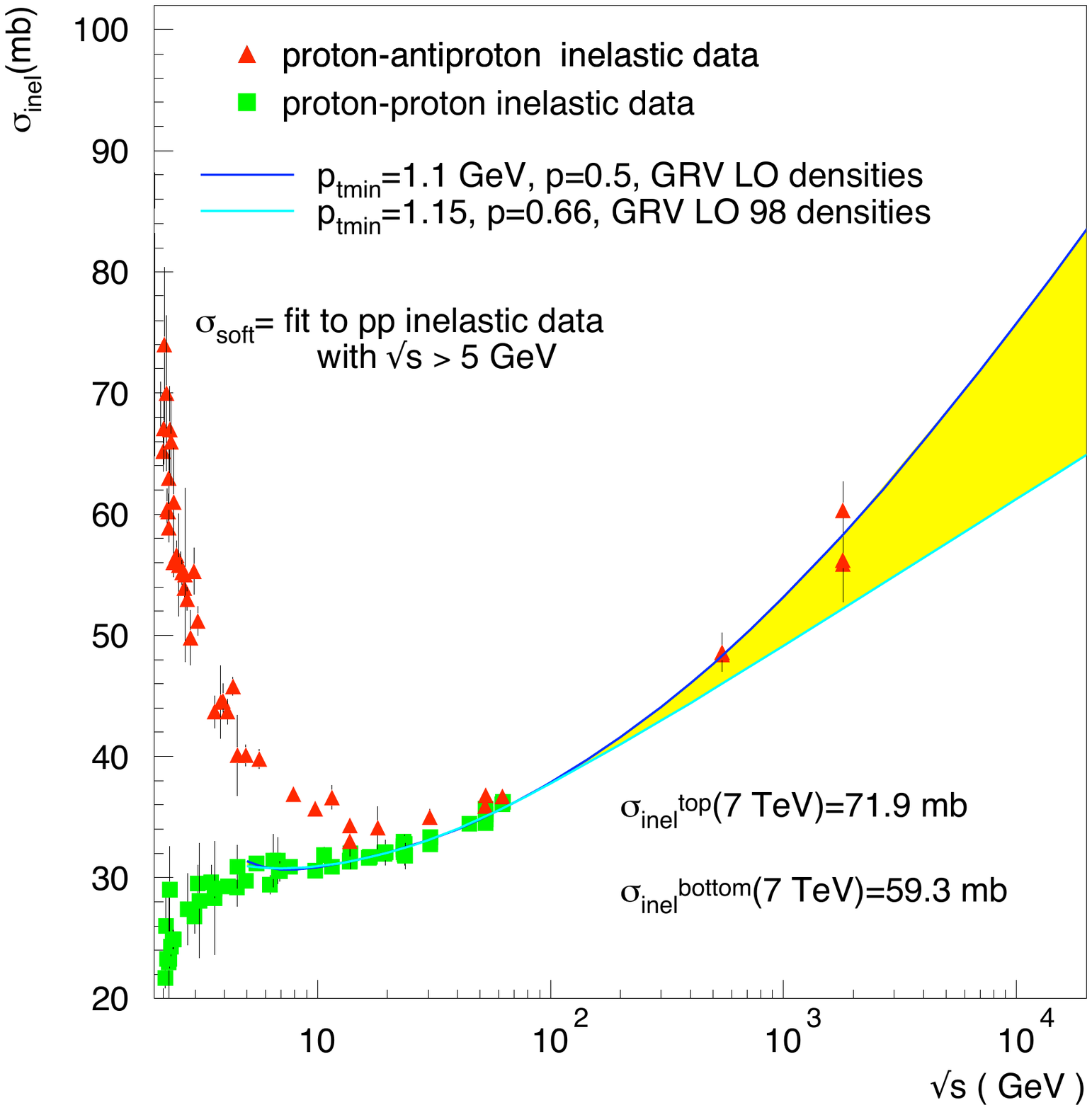}
\end{minipage}
\vspace{-2cm}
\caption{Predictions  for the inelastic \pp \ cross-section from  model II-A (left) and Model II-O  (right).}
\label{fig:inelbandagneseolga}
\end{figure}

 We combine the  results of   Fig.~\ref{fig:inelbandagneseolga} in Fig.~\ref{fig:inelbands}.}  In this figure,  using both Model II-A and Model II-O to constrain the low energy behaviour,  we indicate  all the parameter sets used to produce the bands, as well as the numerical values of interest.  We see that  the low energy behaviour is of course better described by Model II-O, which fits low-energy inelastic data, including mini-jets in the overall fit. However, a  good description is also obtained from the same low energy eikonal entering the total cross-section (Model II-A),    an indication  that the two-channel eikonal model works at low energy.
  In this figure, for simplicity,  we  show the application of  Model II-O only for GRV densities. Also, to compare our choice of high energy parameters   with those used for the total cross-section, we have plotted a  curve { (black line)} with same $p_{tmin}=1.15 \ GeV$ and GRV densities  as in Fig.~\ref{fig:september24-tot-inel-GRV},  but with $p=0.5$, a value  which we find to give a good  description of the pre-LHC inelastic data.


\section{Elastic, inelastic and diffraction processes in eikonal models\label{sec:comment}}
\begin{figure}
\centering
\vspace{-5cm}
\resizebox{1\textwidth}{!}{
\includegraphics{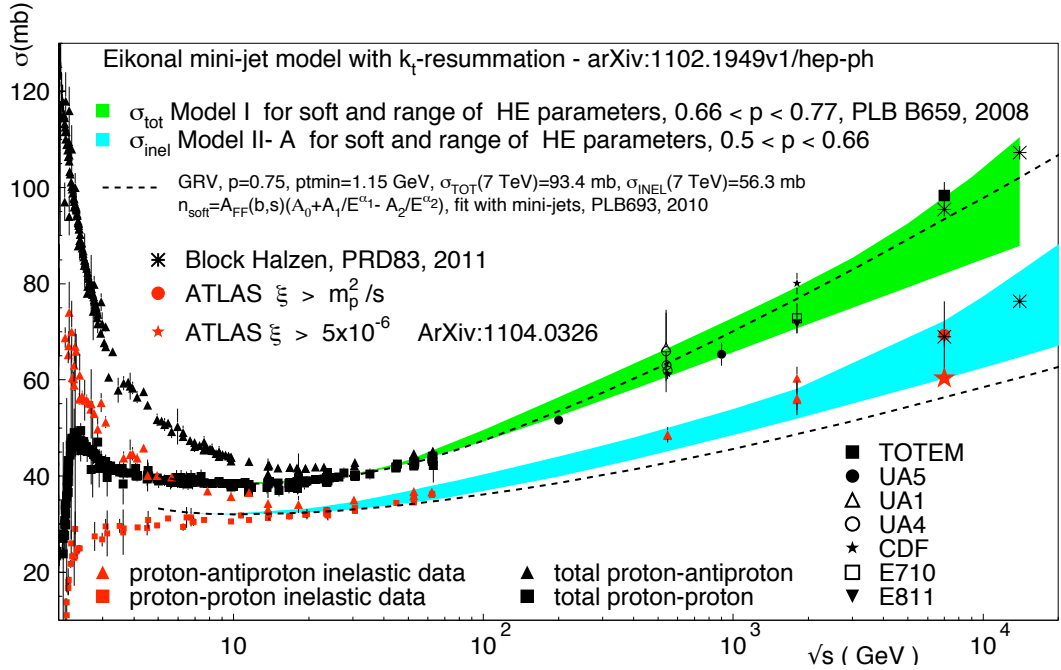}
}
\vspace{-8cm}
\caption{Total and inelastic data for $pp$ and ${\bar p}p$ scattering compared 
with the prediction from our model (dashed curve)\cite{Achilli:2011sw}, in the two-channel case, when the same eikonal function is used both for the total and the inelastic cross-sections, from low to high energy. The upper band is the same as in Fig.\ref{fig:total7tev}, the lower   band  for the inelastic cross-section is the same as  in the left panel of  Fig.~\ref{fig:inelbandagneseolga}. Comparison is made with   theoretical predictions by Block and Halzen \cite{Block:2011uy}, with  TOTEM data \cite{Latino:arXiv1110.1008} for the total cross-section, ATLAS  inelastic data extrapolated  to $\xi > m^2_p/s$ and with ATLAS data in the range 
 $5\times 10^{-6}$, where $\xi=M^2_X/s$ (star symbol) \cite{
 Aad:2011eu}.}
\label{fig:alltogether}
\end{figure}

In the previous section, we have seen that, up to TeVatron energies,  the  two channel model cannot  
 accomodate both inelastic (diffraction included)  and total cross-section data without a change of parameters. On the other hand  by changing the singularity parameter $p$, in a way consistent with our understanding of the role played by the parameter, 
the  model results can span  both the total and the inelastic data at energies up to the TeVatron (blue band). We shall now 
 combine the results of our eikonal mini-jet model for the total and the
inelastic cross-section in a single figure and compare the latter with   ATLAS recent results \cite{
Aad:2011eu} and with 
a recent calculation by
Block and Halzen \cite{Block:2011uy}.  In
Fig.~\ref{fig:alltogether}  the upper  (green) band is from  Fig.~\ref{fig:total7tev} and the lower (blue)  band from Fig.~\ref{fig:inelbands} .
  Once a set of parameters has been fixed from the total cross-section data, the dashed curves   indicate  what the  two-channel eikonal mini-jet model would predict, without changing any parameter in going from the total to the inelastic. As already mentioned, the dashed  curve for the
inelastic cross-section does not fit the TeVatron data, nor the $S{\bar p}pS$  
data, but its comparison with   the recently released  ATLAS measurement   or preliminary  CMS data in the central region can be  very instructive. The dashed line  (see Fig.~\ref{fig:september24-tot-inel-GRV}), corresponding to the model with the same $p$ value  as in the total cross-section curves,
 is close
 to the preliminary value reported by the CMS Collaboration for inelastic events with  at least 2 charged particles
 with $\eta<2.4$  and $p_{\perp}>200 MeV/c^2$~\cite{cmsnote}
This confirms the interpretation of the two-channel eikonal minijet model, 
put forward in Sect. II, namely that this model describes total inelastic 
processes, with little or no correlations.
 
 On the other hand, with the present parametrization, but changing the parameter $p$,  we  find 
  that our lower (blue)  band  is  consistent with the ATLAS results, as corrected for diffraction \cite{Aad:2011eu} (full circle).

 Let us now discuss the missing element in Eq.~(\ref{eq:inex1}), that is a
 category of processes which are inelastic but correlated, namely single and double diffraction. In  diffraction, partons in the final state resulting from multiple collisions or from soft gluon emission must recombine into a proton or are constrained to be emitted   along the outgoing proton. It would then  be more correct to write Eq.( \ref{eq:inex2}) as
\begin{equation}
\label{eq:inex3}
\sigma_{inel}^{uncorrelated}(s)=\sum_{n=1}\int d^2{\bf b} \ P(\{n,\nbar\})=\int d^2{\bf b}[1-e^{-\nbar (b,s)}]
\end{equation}  
so that, generally speaking,  
\begin{equation}
\sigma_{total}=\sigma_{elastic}+\sigma_{inel}^{diff}+\sigma_{inel}^{uncorrelated}
\end{equation}
and include in $\sigma_{inel}^{diff}$ all correlated processes, like single and double diffraction.
Thus Eq. ~(\ref{eq:inex1}) can be used only for inelastic uncorrelated processes, while  the inelastic  correlated processes are to be described separately. 

On the other hand,  
the above considerations do not yet suggest how to include the diffractive processes. The solution adopted in this paper, illustrated by the lower (blue) band in Fig.~\ref{fig:alltogether}, has been to mimick the presence of diffraction processes using  different values for  the singularity parameter $p$ in Eqs.~ (\ref{eq:inex2}) and (\ref{eq:1}).
 By comparing with previous data,  Figs.~\ref{fig:inelbands} and \ref{fig:alltogether} indicate that  the value   $p=0.5$, with the present LO parameterization (LO GRV and MRST),  gives the best description of  the total inelastic cross-section, while the above  discussion, following the one in in Sec.~\ref{sec:ineltotal}, indicates that values  with $p=0.75$ would only include uncorrelated inelastic processes. 

 For the future,  once the parameters of the model have been tuned to the measured total cross-section, more precise predictions for the totality of uncorrelated inelastic events could  be extracted from our model.

\subsection{Uncertainties and the nature of the theoretical band\label{subsec:band}}
For clarity and completeness, we recapitulate here the uncertainties in our model and the nature of the bands we have presented.
The  bands  in this paper preceded the release of LHC data for total and inelastic cross-sections, and reflect how well the model worked in describing lower energy data \cite{Godbole:2008hx,Achilli:2011sw} as well as their  uncertainties, mostly  the TeVatron data for  the total cross-section. For the future, the new data will allow a better tuning of all the parameters involved and hence firmer predictions for total cross-section data at ultra high energies or for quantities, such as  survival probabilities,  needing the inelastc cross-section.

Even so, the bands we have presented  are in reasonable agreement with  the new data and provide confirmation of our understanding of the eikonal two-channel model. The blue (lower ) band in Fig.~ \ref{fig:alltogether} describes the interplay between all the HE energy parameters. 
 The essential parameters are the PDFs, $p_{tmin }
$ and $p$.   The   PDFs control both the rise (mini-jets) as well as the logarithmic asymptotic behavior  through $q_{max}$. At fixed $p_{tmin}$, different PDF's lead to different values for $\sigma_{jet}$ and for $q_{max}$.  In the present  formulation, $\sigma_{jet}$ are controlled by uncertainties in the LO low-$x$ behaviour of the gluon densities, and $q_{max}$  by the LO valence quark densities.  The lower is $p_{tmin}$ the faster the rise, and, at high energies, the  more this rise needs to be quenched. The quenching of the rise depends on how much is the allowed loss in transverse momentum, $q_{max}$ and upon  the degree of singularity of the effective coupling, the parameter $p$,
  for which we  have both an upper as well as a lower bound. For the convergence of the IR integrals, it is incumbent that $p<1$. [It must be recalled that the Wilson equal area rule leading to a strictly linear potential viz., $p=1$ is not allowed in our  resummed  model]. On the other hand,
it is easy to check that there is no confining potential if $p<1/2$. 

Within the contraint $1/2 <p<1$, values for this parameter are determined through the phenomenological description of the data.
There is an interesting interplay between the phenomenological value of $p$ and its weight in describing the uncorrelated versus correlated events. In the model, once the value for $p$ has been  fixed through the total cross-section, it can be used to describes the uncorrelated inelastic events, while the fully inelastic data need a different treatment. We have mimicked the amount of diffraction to include in the full inelastic cross-section by decreasing values of $p$. 
As $p$ is decreased, the amount of acollinearity introduced by the soft-gluon radiation changes. We see, that a description of the underlying fully inelastic data   would favour  $p = 0.5$. 

Finally, we point out that some uncertainties are also introduced by the different purely phenomenological models for the quantity ${\bar n}_{soft}$ and are reflected in the band for the inelastic cross-section.

\section{Conclusions}

 We have presented  results from  the two channel eikonal model for the inelastic cross-section. We have shown that such formulation  only describes uncorrelated processes, a  result  independent of the details of the model used to describe the data. This observation clarifies the origin of a long standing difficulty  of the two channel eikonal models well known to people working in total and inelastic cross-sections. 

We have  shown numerical  results 
for the  description of  total and  inelastic  cross-sections at LHC at $\sqrt{s}=7\ TeV$ and beyond, using our QCD based model. Our numerical results are obtained in a LO QCD model for collisions, embedded in the eikonal formulation. The major QCD phenomenological input is from Parton Density Functions, which can be   considered  QCD  parameters  through which we build our eikonal in order to describe the two high energy effects: the rise at the beginning and the slowing down towards  asymptotia. While  it is difficult to describe these two effects in QCD models which satisfy unitarity and the Froissart bound, in our model we have both.

We have found  that  with exactly the same low energy  parameters entering the 
total cross-section calculation, but with different values of the high energy
parameters $p$ and $\ptmin$, our model predictions catch the  TeVatron data as 
well as new data from the ATLAS collaboration \cite{Aad:2011eu}.  On the other 
hand, if no parameters are changed at all in the eikonal function used for 
$\sigma_{total}$, 
our results agree with the hypothesis that the simple two-channel eikonal mini-jet models describe only the non-diffractive part of the inelastic cross-section. We defer to future work an analysis of the inelastic cross-section which might include  correlations in the diffractive terms, or   inclusion of a non zero $\Re e \chi(b,s)$.

\section*{Acknowledgements}
We acknowledge conversations and discussions  with L. Anchordoqui and thank 
him for providing us with the numbers we have used for the inelastic 
cross-sections. G.P. thanks the MIT Center for Theoretical Physics and  
Brown University Physics Department for hospitality while this work was 
being written. Y.N.S. thanks Northeastern University Physics Department 
for hospitality.   Work partially supported by 
Spanish MEC (FPA2006-05294, FPA2010-16696 and ACI2009-1055) and by
Junta de Andalucia (FQM 101). This work has been supported in part by MEC (Spain) under Grant
FPA2007-60323 and by the Spanish Consolider Ingenio 2010 Programme CPAN
(CSD2007-00042).
R.M.G. wishes to acknowledge support from the Department of Science and 
Technology, India under Grant No. SR/S2/JCB-64/2007, under the J.C. Bose
Fellowship scheme. She also wishes to thank the theory division at CERN 
during the preparation of an earlier  version of this manuscript.
 \appendix
\section{The average  number of soft collisions ${\bar n}_{soft}(b,s)$}
\label{sec:appendix}%
As in the case of the hard collisions,  we approximate $\nbar_{soft}(b,s)$ with the factorized expression
\begin{equation}
\nbar_{soft}(b,s)=A_{soft}(b,s)\sigma_{soft}(s)
\end{equation}
and then use two different  phenomenological parametrizations  for the average number of soft collisions. 
 In the first model, Model I, we assume that the decrease in the proton proton total cross-section is due 
only to soft gluon emission. 
\subsubsection{Model I,  $A_{soft} (b,s)$ given by resummation expression:}
This is the model adopted in our analyses in 
Refs. ~\cite{Godbole:2004kx,Achilli:2007pn}.
 In this case, for the $b$-distribution, we use { Eq.~(\ref{hbs}), i.e}. the BN model. In this equation,  
we parametrize   $q_{max}$ to make it 
vary   slowly with energy, thus
giving  a  slight (decreasing) energy dependence 
to $A_{soft} (b,s)$ \cite
{Godbole:2004kx}.  In Table ~\ref{qmaxtable} from \cite{Godbole:2004kx} we reproduce the values of $q_{max}^{soft}$ used to evaluate $A_{soft}(b,s)$. These values correspond to the observation  that for  processes contributing to $n_{soft}$, 
a soft gluon will always carry away less energy than for those 
contributing to 
$n_{hard}$. The question is how much lower is the allowed energy. The above set of values is purely phenomenological and the values are  those we found to  give an acceptable description for 
$\sigma_{pp}$ before the rise.  
\begin{table}
\caption{\label{qmaxtable} Average $q_{max}$
 values used for the impact parameter distribution of the soft part of 
the eikonal}
\begin{tabular}{|c|c|}
\hline
$\sqrt{s}(GeV)$&$q_{max}^{soft}(GeV)$\\
5.&0.19\\
6.&0.21\\
7.&0.22\\
8.&0.23\\
9.&0.235\\
10.&0.24\\
50.&0.24\\
100&0.24\\ \hline
\end{tabular}
\end{table}

For $pp$ scattering, where the s-channel is exotic and no trajectories are 
exchanged in the $t-$channel, our use of soft gluon emission  using
a form for $A_{BN}(b,s)$  a-la Eq.~(\ref{hbs}) with $q_{max}$ chosen as 
explained above, is sufficient to explain the 
slight decrease at low energies, in the total $pp$ cross-section. The situation 
is more complicated for \pbarp. 
As described in ~\cite{Godbole:2004kx}, 
we then simply write for the cross-section
\begin{equation}
\sigma_{soft}(s)=\sigma_0
[1+\frac{2 \epsilon}
{\sqrt{s}}
]
\end{equation}
with  $\sigma_0$ a constant, as in Table~ \ref{tab:andrea} and $\epsilon=0,1$ for $pp$ and \pbarp \ respectively.  
\subsubsection{ Model II, $A_{soft} (b,s) $ given by $A_{FF} (b,s)$: } 
In the second approach,   for the 
soft collisions we have used the Form Factor  model for $A(b,s)$, Eq.(\ref{eq:aff}),together with a standard, Regge exchange inspired 
parametrization of  \pbarp \  and $pp$ cross-sections 
at low energy, namely
\begin{equation}
\sigma_{soft}(s)=\sigma_0+\frac{A_1}{(E_{lab})^{\alpha_1}} \mp \frac{A_2}{(E_{lab})^{\alpha_2}}
\label{eq:sigsoftparam}
\end{equation}
with $\mp$ referring to $pp$ and $ {\bar p}p$ respectively,  and the constant parameters determined 
through a  fit to the low energy data. In earlier publications, 
such as Ref.~\cite{Grau:1999em}, the fit 
to low energy data was done using an expression which did not  include the 
mini-jet contribution, whereas in our more recent analysis~\cite{Grau:2010ju} 
the minijet contribution  is included in the fit.  From the point of view of the $\chi^2$,
the difference in the results is not noticeable.   In Table~\ref{tab:softparam} from  \cite{Grau:2010ju} we reproduce the values of various  parameters used in Eq. ~(\ref{eq:sigsoftparam}).

\begin{table*}
\caption{Results of the fit to 
\pp \ and \pbarp \ data.}
\label{tab:softparam}
\centering
  \begin{tabular}
{
|p{7cm}|p{7cm}|}
\hline
Fit  for \pp
& Fit  for \pbarp       \\ 
\hline
 $\sigma_0 = (48.20 \pm 0.19)$ mb& $\sigma_0 = (47.86 \pm 2.47)$ mb \\
$A_1 = 101.66 \pm 16.35$  \hspace{0.2cm}
$\alpha_1 = 0.99 \pm 0.13$       &$A_1 = 132.07 \pm 32.89 $  \hspace{0.5cm}
$\alpha_1 = 0.69 \pm 0.14$ \\
$A_2 = 27.89 \pm 4.78$ \hspace{0.2cm} $\alpha_2 = 0.59 \pm 0.04$
      &  $A_2 = 0.82 \pm 0.31 $  \hspace{0.5cm} $\alpha_2 = 0.52 \pm 0.07$
       \\  $\chi^2 = 154.1/(102+5-1)$ & $\chi^2=24.65/(31+5-1)$\\
   \hline
     \end{tabular}
\end{table*}

\bibliography{lhc7}
\end{document}